\documentclass[preprint,12pt]{elsarticle}

\usepackage{epsfig,graphicx,float,pst-all}
\usepackage{mathrsfs}
\usepackage[utf8]{inputenc}
\usepackage{rotating}
\usepackage{url}
\usepackage{amsmath}
\usepackage{amssymb}
\usepackage{color}

\usepackage{natbib}
\usepackage{hyperref}
\usepackage[warn]{textcomp}
\usepackage{gensymb}
\usepackage{footnote}
\usepackage{siunitx}
\usepackage{comment}
\usepackage{lipsum}
\usepackage[normalem]{ulem}

\journal{Nuclear Physics A}

\begin{document}
	
	\begin{frontmatter}
		
		
		\title{Exploring the structure of $^{29}$Ne}

			
		\author[add1]{Manju}\corref{mycorrespondingauthor}
		\cortext[mycorrespondingauthor]{Corresponding author}
		\ead{manju@ph.iitr.ac.in}
		
		\author[add1]{M. Dan}
		\ead{mdan@ph.iitr.ac.in}
		
		\author[add2,add3]{G. Singh}
		\ead{gsingh@us.es}
		
		\author[add4,add5]{Jagjit Singh}
		\ead{jsingh@rcnp.osaka-u.ac.jp}
		
		\author[add6]{Shubhchintak}
		\ead{shubhchintak@ulb.ac.be}
		
		\author[add1]{R. Chatterjee}
		\ead{rchatterjee@ph.iitr.ac.in}
		
		\address[add1]{Department of Physics, Indian Institute of Technology - Roorkee, 247667, India}
		\address[add2]{Departamento de F\'{i}sica At\'{o}mica, Molecular y Nuclear, Facultad de F\'{i}sica, Universidad de Sevilla, Apartado 1065, E-41080 Sevilla, Spain} 
		\address[add3]{Dipartimento di Fisica e Astronomia``G.Galilei", Università degli Studi di Padova, via Marzolo 8, Padova, I-35131, Italy}
		\address[add4]{Research Centre for Nuclear Physics (RCNP), Osaka University, Ibaraki 567-0047, Japan}
		\address[add5]{Department of Physics, Akal University,Talwandi Sabo, Punjab, India-151302}
		\address[add6]{Physique Nucl\'{e}aire Th\'{e}orique et Physique Math\'{e}matique, C. P. 229, Universit\'{e} Libre de Bruxelles (ULB), B 1050 Brussels, Belgium}
		
		\begin{abstract}
		We apply a fully quantum mechanical Coulomb breakup theory under the aegis of post form finite-range distorted wave Born approximation to analyze the elastic Coulomb breakup of $^{29}$Ne on $^{208}$Pb at $244$\,MeV/u. We calculate several reaction observables to quantify its structural parameters. One-neutron removal cross section is calculated to check the consistency of the ground state configuration of $^{29}$Ne with the available experimental data. A scrutiny of the parallel momentum distribution of the charged fragment reveals a full width at half maximum of $82$\,MeV/c, which is in good agreement with the experimental value and indicates a moderate halo for a nearly spherical $^{29}$Ne in the $^{28}$Ne$(0^+) \otimes 2p_{3/2}\nu$ ground state. The energy-angular distributions and average momentum of the charged fragment point to the absence of post-acceleration effects in the breakup process, a desirable result for elastic breakup.
		\end{abstract}
		

\end{frontmatter}


\section {Introduction}\label{intro}

Sophisticated upgrades in the radioactive ion beam facilities all over the world have led to prodigious advancements in the understanding of nuclear shell evolution while moving away from the line of stability. One small region in the neutron rich side of the nuclear chart around neutron number \textit{N} $\approx20$, the so called ``island of inversion" \citep{War1990}, exhibits interesting constitutional properties, such as quenching of the conventional shell gaps \cite{Sorlin2008}, exotic halo structures and deformation \cite{War1990, moto95}. The ground states of the residents of this island, whose major portion is spanned by the medium mass neutron rich isotopes of Ne, Na and Mg, are dominated by the mixing of $fp$-shell `intruder' and the conventional $sd$-shell neutron configurations, leading to the melting of the $N\approx20$ shell closure. Recently, this trend has also been seen in fluorine isotopes ($^{29}$F and $^{31}$F) which has triggered off several investigations \cite{Singh2020, Fortunato2020, Michel2020, Masui2020}.

Ever since the first foot-prints of the $N\approx20$ shell softening were found by mass measurements in a Na isotope \citep{THI75}, considerable efforts have been devoted to track down the limits of the island of inversion and the proliferation of the intruder strengths along an isotopic chain. In the Ne isotopic chain, various experimental studies \cite{Yana2003, DOOR16, Door2009, Nakamura2014, Terry2006, Koba2016} confirm the genesis of intruder configurations in $^{28}$Ne. Further, $^{29-32}$Ne are located well inside this island, providing exciting contenders for research in this area. In fact, both experimental and theoretical studies for $^{31}$Ne have confirmed it as a halo nucleus with intruder neutron configuration \cite{Nakamura2014,Nakamura2009,hourichi2010,Shubh2014,hong17}.

For the present study, we focus on the medium mass nucleus $^{29}$Ne (with $Z=10$ and $N=19$), 
which is a loosely-bound system with the measured value for the one-neutron separation energy $(S_{n})$ equal to $0.96(0.14)$\,MeV \cite{AUDI12, Jurado2007}.
In the conventional shell model picture, the valence neutron in $^{29}$Ne is expected to occupy the $1d_{3/2}$-orbital, leading to a ground state spin-parity of ${3/2}^+$. Indeed, the $\beta$-decay data on $^{29}$Ne was analyzed on the basis of this assumption \cite{Vandana2006}. However, there are two scenarios which rule out this spherical 
shell model picture: the consideration of the lowering of the intruder $fp$-shell, and the elevation of the $2s_{1/2}$-orbital \cite{MISU97, Hamamoto2004}.
The feasibility of the first case was boosted by Monte Carlo shell model calculations \cite{Utsuno1999, Koba2016}, 
where it was claimed that the ${3/2}^+ (1d_{3/2})$ and the ${3/2}^- (2p_{3/2})$ states are substantially degenerate, with 
an energy gap of the order of $73$\,KeV. A chronological experimental study conducted 
for the interaction cross section measurements of Ne isotopes recorded a significant difference in the cross sections for $^{28}$Ne and
$^{29}$Ne, suggesting low $l$, i.e., $s$-dominance in the ground state of $^{29}$Ne and thus, promoting the second possibility \cite{Tak2012}.
Similarly, the one-neutron ($1n$) knockout measurements from $^{30}$Ne have reported narrow inclusive momentum distributions, a further indication of the dominance of large $2p_{3/2}$ and/or $2s_{1/2}$ orbitals and small $1d_{3/2}$ orbital\cite{Liu2017}.
Be that as it may, all these assessments do not significantly constrain the spin-parity assignment for the ground state of $^{29}$Ne.

Recently, a more conclusive assignment of the spin-parity was made in the enhanced one-neutron removal cross section measurements on C and Pb targets at $240$\,MeV/u and $244$\,MeV/u, respectively, through a combined investigation of Coulomb and nuclear breakup \cite{Koba2016}. 
The ground state spin-parity $(J^{\pi})$ of $^{29}$Ne was predicted to be $3/2^{-}$ with a low-lying excited state $3/2^{+}$. The low-lying $7/2^{-}$ state that should ideally accompany the $p$-subshell in case of the lowering of the $fp$-shell was not observed. The measurement, thus, concluded a deformation driven `moderate' $p$-wave halo structure for $^{29}$Ne in its ground state. This new experimental information provides a stringent test for theory and sheds new light on the understanding of the island of inversion.

Therefore, in view of these developments, we conduct the first theoretical study of the elastic Coulomb breakup of $^{29}$Ne on $^{208}$Pb target at $244$\,MeV/u within the post form finite-range distorted wave Born approximation (FRDWBA) theory \cite{Shubh2014,Chatterjee2018}. We apply this theory to evaluate several reaction observables, like relative energy spectra, parallel momentum distribution, angular distributions, etc., to examine the ground state structure of $^{29}$Ne. We also take into account the projectile deformation effects to estimate the quadrupole deformation value ($\beta_2$).
Our post from FRDWBA inherently contains the target-projectile electromagnetic interaction to all orders. Full ground state wave function of the projectile, of any angular momentum configuration, enters as the only input to this theory in order to calculate the transition matrix. The theory also includes the contribution corresponding to all the multipoles and the relative orbital angular momenta between the breakup fragments, from the entire non-resonant continuum.


The paper is organized as follows.
Section~\ref{for} briefly describes the formalism where we discuss the FRDWBA theory of breakup reactions including deformation effects in the projectile ground state. The results of our structure calculations for $^{29}$Ne are presented in Section~\ref{rnd}. Section~\ref{con} conveys the conclusions.
\begin{figure}[h!]
	\centering
	\includegraphics[scale=0.35]{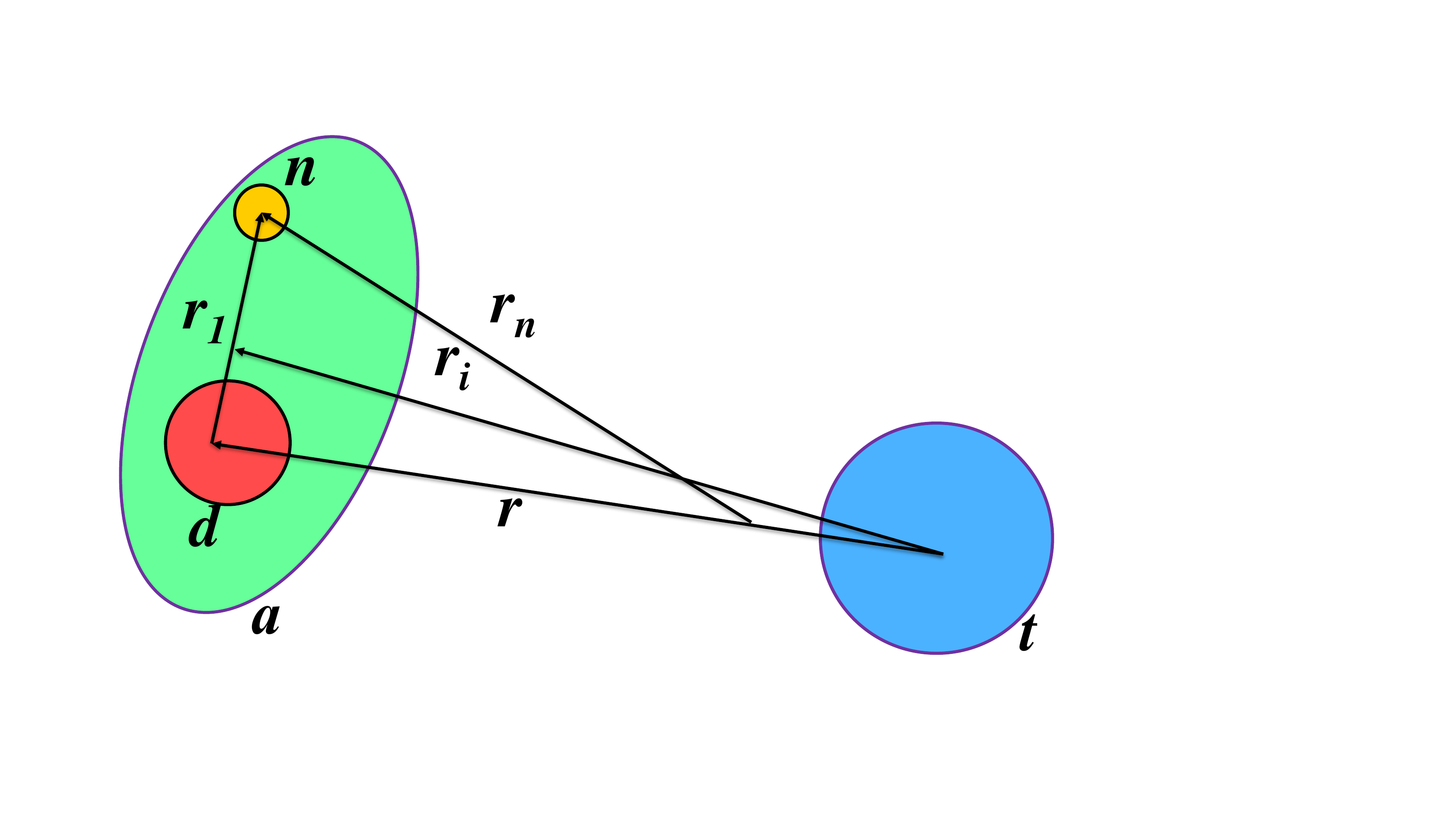}
	\caption{The deformed projectile `\textit{a}' interacting with the target `\textit{t}' in a three-body jacobi coordinate system with positions vectors, \textbf{r} = \textbf{r}$_{\textbf{i}}- \alpha$ \textbf{r}$_{\textbf{1}}$, $\alpha$ = $m_n/(m_n+m_d)$ and \textbf{r}$_{\textbf{n}}$ = $\gamma$\textbf{r}$_{\textbf{1}}$ + $\delta$\textbf{r}$_{\textbf{i}}$, $\delta$ = $m_t/(m_d+m_t)$, $\gamma=1-\alpha \delta$.}
	\label{cors}
\end{figure}

\section{Formalism}
	\label{for}
	
Consider a beam of $^{29}$Ne incident on a $^{208}$Pb target at $244$ \,MeV/u energy. Under the influence of its strong Coulomb field, $^{29}$Ne breaks up elastically into two constituents, $^{28}$Ne and a neutron, i.e., the reaction $^{29}$Ne + $^{208}$Pb $\longrightarrow$ $^{28}$Ne + n + $^{208}$ Pb.
The triple differential cross section for this reaction is given by

\begin{eqnarray}
\dfrac{d^3\sigma}{dE_d d\Omega_d d\Omega_n} = \dfrac{2 \pi}{\hbar \upsilon_{at}} \rho(E_d, \Omega_d, \Omega_n) \sum_{\ell m} \dfrac{1}{(2 \ell + 1)} \vert \beta_{\ell m} \vert^2, \nonumber \\
\label{eqn1}
\end{eqnarray} 

with $a$, $d$, $n$, and $t$ being the projectile ($^{29}$Ne), core ($^{28}$Ne), valence neutron ($n$), and the target ($^{208}$Pb), respectively. $v_{at}$ in Eq.(\ref{eqn1}) is the $a-t$ relative velocity in the initial channel while $\rho$ is the appropriate three-body phase space factor \citep{Fuchs1982}. The reduced transition amplitude in the post form FRDWBA is given by

\begin{eqnarray}
\hat{\ell}\beta_{\ell m}=\int d\textbf{r}_{i} e^{-i \delta \textbf{q}_n.\textbf{r}_i} \chi_{d}^{(-)*}(\textbf{q}_d, \textbf{r}_i) \chi_{a}^{(+)}(\textbf{q}_a, \textbf{r}_i)\nonumber\\
\times  \int d\textbf{r}_1 e^{-i\textbf{Q}.\textbf{r}_1}V_{dn}(\textbf{r}_1)\phi_a^{\ell m}(\textbf{r}_1),
\label{eqn2}
\end{eqnarray}

where $Q=\gamma \textbf{q}_n-\alpha \textbf{K}$. In Eq. (\ref{eqn2}), $\textbf{q}_j$ are the Jacobi wave vectors corresponding to the position vectors $\textbf{r}_j$ as shown in Fig.~\ref{cors} and $\textbf{K}$ is the effective local momentum of core-target relative system whose direction is same as that of $\textbf{q}_d$. $\alpha$, $\gamma$ and $\delta$ are the mass factors, while $\chi$'s are the pure Coulomb distorted waves with incoming (-) and outgoing (+) wave boundary conditions. $\phi_a^{\ell m}(\textbf{r}_1)$ represents the ground state wave function of the projectile $a$ and is the only input to this Coulomb dissociation theory. 

The integral over $\textbf{r}_i$ in Eq.(\ref{eqn2}) represents the dynamics of the reaction and can be analytically expressed in terms of what is known as the `Bremsstrahlung integral' \citep{Nord1954}. The second integral over $\textbf{r}_1$ highlights the structure part as it contains information about the anatomy of the projectile nucleus where systemic effects like deformation are introduced via the interaction potential $V_{dn}(\textbf{r}_1)$. For our case study, we take $V_{dn}(\textbf{r}_1)$ to be of axially symmetric quadrupole deformed Woods-Saxon type \footnote{The quadrupole deformed axially-symmetric Woods-Saxon potential is
	\begin{eqnarray}
	V_{core+n}&=&V^{0}_{ws}\Bigg(1-\beta_2RY_{2}^{0}(\hat{\textbf{r}}_1)\frac{d}{dr_1}\Bigg)f(r_1),\nonumber\\
	\label{vdn}
	\end{eqnarray} 
	where, $f(r_1)=\Bigg[1+exp\Bigg(\dfrac{r_1-R}{a_0}\Bigg)\Bigg]^{-1}$ and $R=r_0 A^{1/3}$. $V^{0}_{ws}$ is the depth of the Woods-Saxon potential and $\beta_2$ is the deformation. $r_0$ and $a_0$ are the radius and diffuseness parameter.}. Apart from reducing the computational complexity, resolving the reduced transition amplitude into the two component integrals of Eq.(\ref{eqn2}) manifests that any deformation effects will only ensue changes in the structure part, the dynamics part remains unaffected \citep{Chatterjee2000,Shubh2014}. For more details on the FRDWBA formalism, one may refer to \citep{Chatterjee2018,Shubh2014,Shubh2015, Gagan2016}. The results are discussed in the next section.

	\section{Results and discussions}\label{rnd}
\subsection{Structure aspects and ground state wave function of $^{29}$Ne}

$^{29}$Ne has a one-neutron separation energy of $S_n=0.96$ (0.14)\,MeV \citep{AUDI12}. The ground state spin-parity of $^{29}$Ne is not well defined and it can be formed either of $^{28}$Ne$(0^+) \otimes 2s_{1/2}\nu$, $^{28}$Ne$(0^+) \otimes 2p_{3/2}\nu$, $^{28}$Ne$(0^+) \otimes 1d_{3/2}\nu$ or $^{28}$Ne  $(0^+) \otimes 1f_{7/2}\nu$ configurations (see for example discussion in Ref. \citep{Koba2016}). The ground state wave function can be obtained by solving the coupled Schr\"odinger equation with axially symmetric quadrupole deformed potential (Eq.~\ref{vdn}). 
Due to the deformation, for a given $\ell$, the radial wave function will have an admixture of the other $\ell$ components of the same parity. Thus, such wave function should be relatively different from those obtained with spherical Woods-Saxon potential. However, to retain the analytical flavor of our calculations, we have calculated the radial wave function from a spherically symmetric Woods-Saxon potential for a single $\ell$. This point is valid as it has been shown in Ref. \citep{Hamamoto2004}, that if the binding energy of the deformed projectile is not large, the lowest $\ell$-component dominates in the neutron orbits of the deformed potential, irrespective of the magnitude of the deformation. Indeed, we have verified this point in the present case as it was done in Ref. \cite{Shubh2017}, and found that the deformed wave function is not very different from the spherical one for $\beta_2$ up to 0.5.

For a given set of radius ($r_0 = 1.24\,fm$) and diffuseness ($a_0= 0.62\,fm$) parameters (chosen from Ref. \cite{Shubh2014}), the depth of the Woods-Saxon potential is adjusted to reproduce the exact ground state binding energy ($S_n=0.96$\,MeV) of the projectile nucleus. The depths of the Woods-Saxon potential for the above mentioned four possible configurations with respect to the valence neutron are depicted in Table~\ref{tab1}.
\begin{table}[ht]
	\caption{Depths of the Woods-Saxon potential, $V_{ws}^{0}$ corresponding to neutron removal from the $2s_{1/2}$, $2p_{3/2}$, $1d_{3/2}$ and $1f_{7/2}$ orbitals of $^{29}$Ne at one-neutron separation energy, $S_n=0.96$\,MeV. Parameters $r_0$ and $a_0$ as $1.24$\,fm and   $0.62$\,fm, respectively.}  
	\centering 
	\begin{tabular}{c c } 
		\hline\hline 
		\\
		~~Ground state orbitals~~ &  ~~Potential depth ($V_{ws}^{0}$)~~\\ [1ex] 
		\hline 
		$2s_{1/2}$ & 33.8 \\ 
		$2p_{3/2}$ & 54.9 \\
		$1d_{3/2}$ & 38.6 \\
		$1f_{7/2}$ & 52.9 \\ [1ex]
		\hline 
		\hline
	\end{tabular}
	
	\label{tab1} 
\end{table}

\subsection{One-neutron removal cross section}
\label{1n}

$^{29}$Ne lies in the region where melting of $N=20$ and $N=28$ shell closure occurs due to the intruder states from $fp$-shell \citep{War1990}. As $1d_{3/2}$ and $1f_{7/2}$ states move closer towards each other the shell gap at $N=20$ breaks down \citep{Sorlin2008}, which suggests that contributions from $f_{7/2}$ configuration are expected at low excitation energies and even form a significant non-negligible part of the ground state (g.s.) wave function of nuclei in this region. Shell-model calculation predicted four low-lying energy states of $^{29}$Ne, with spin parities $3/2^+$, $3/2^-$, $7/2^-$, and $1/2^+$ \footnote{Among all these possibilities the states $3/2^+$ and $3/2^-$ are nearly degenerate.} \citep{Koba2016}. Since there are uncertainties in the ground states spin-parity of $^{29}$Ne, therefore, in the present case, we examine the one-neutron removal from the $2s_{1/2}$, $2p_{3/2}$, $1d_{3/2}$ and $1f_{7/2}$ orbitals that can possibly contribute to the ground state of $^{29}$Ne resulting in the above mentioned four possible spin-parities. Despite this variability in its ground state configuration, $^{29}$Ne has a well known one-neutron separation energy, $S_n=0.96(0.14)$\,MeV \citep{AUDI12, Jurado2007}, which will be used for all calculations discussed in this manuscript, unless specified.

\begin{figure}[h!]
	\centering
	\resizebox{0.80\textwidth}{!}{\includegraphics[clip,trim=0cm 0cm 0cm 0cm]{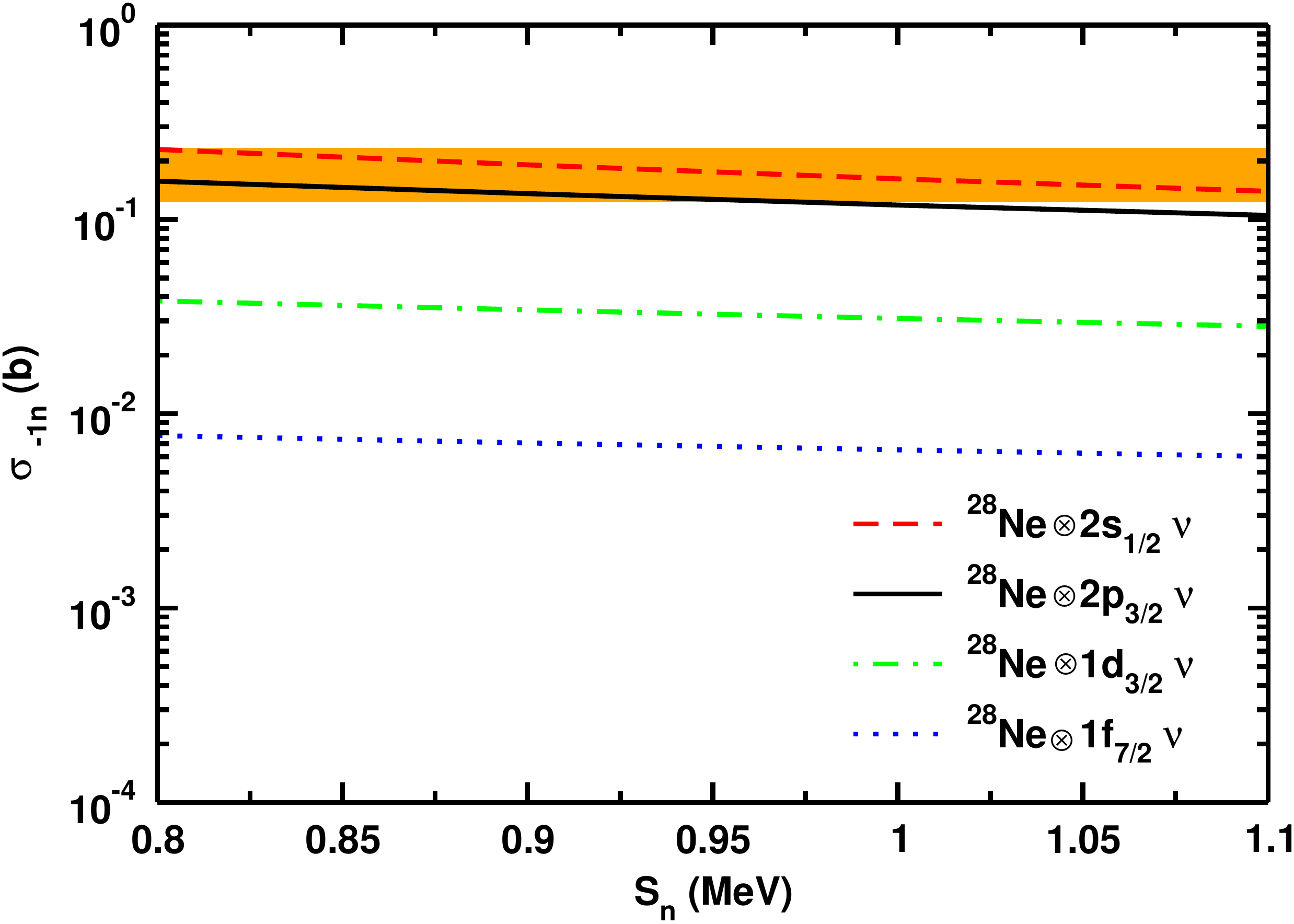}}
	\caption{One-neutron removal cross sections, $\sigma_{-1n}$, in the breakup of $^{29}$Ne on a Pb target at $244$\,MeV/u beam energy as a function of one-neutron separation energy, $S_n$ within the error bars [0.96$\pm$0.14 MeV]. The curves are obtained with different g.s. configurations $^{28}$Ne$(0^+) \otimes 2s_{1/2}\nu$ (dashed line), $^{28}$Ne$(0^+) \otimes 2p_{3/2}\nu$ (solid line), $^{28}$Ne$(0^+) \otimes 1d_{3/2}\nu$ (dot-dashed line) and $^{28}$Ne$(0^+)\ \otimes 1f_{7/2}\nu$ (dotted line) for $^{29}$Ne, using a spectroscopic factor of $1.0$ in each case. The experimental cross section (taken from Ref \cite{Koba2016}) is shown by the shaded band.}
	\label{xs}
\end{figure}

\begin{figure}[h!]
	\centering
	\resizebox{0.80\textwidth}{!}{\includegraphics[clip,trim=0cm 0cm 0cm 0cm]{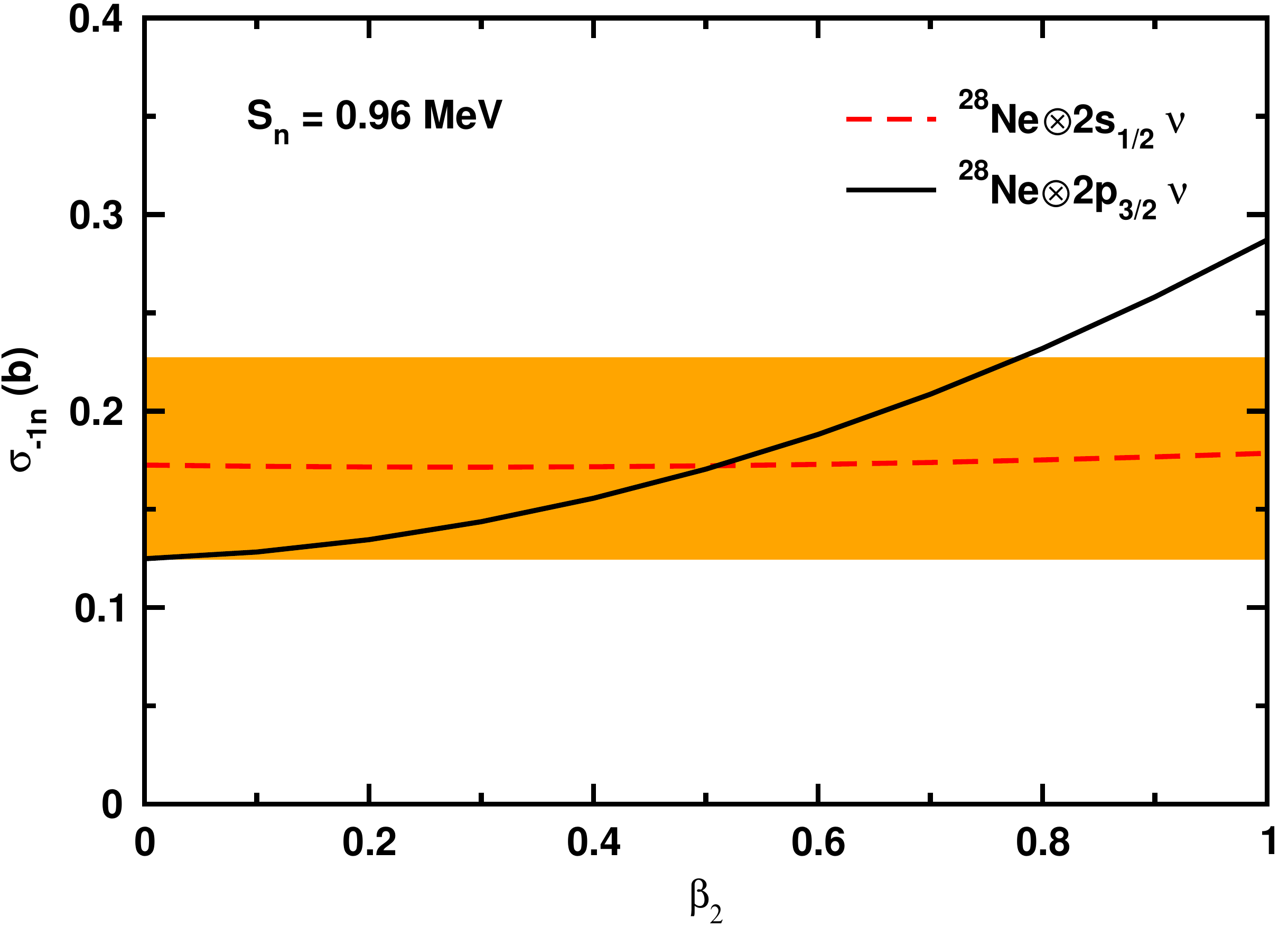}}
	\caption{One-neutron removal cross sections, $\sigma_{-1n}$, as a function of the deformation parameter $\beta_2$ in the Coulomb breakup of $^{29}$Ne on a Pb target at $244$\,MeV/u beam energy for the configurations $^{28}$Ne$(0^+) \otimes2s_{1/2}\nu$ (dashed line) and $^{28}$Ne$(0^+) \otimes2p_{3/2}\nu$ (solid line). The experimental data \cite{Koba2016} are shown by the shaded band.}
	\label{1nx}
\end{figure}

We first investigate the possible ground state configurations which can reproduce the experimental one-neutron removal corss section \cite{Koba2016} in the given separation energy range. In Fig.~\ref{xs}, we present the total one-neutron removal cross section, $\sigma_{-1n}$, for the elastic Coulomb breakup of a spherical $^{29}$Ne on $^{208}$Pb at $244$\,MeV/u, as a function of $S_n$ for different possible ground state configurations: $^{28}$Ne($0^+$)$ \otimes 2s_{1/2}\nu$, $^{28}$Ne($0^+$)$ \otimes 2p_{3/2}\nu$, $^{28}$Ne($0^+$)$ \otimes 1d_{3/2}\nu$ and $^{28}$Ne($0^+$)$ \otimes 1f_{7/2}\nu$. 
The spectroscopic factor ($C^2S$) is assumed to be $1.0$ in each case. 
The shaded band in Fig.~\ref{xs} illustrates the experimental cross section from Ref. \citep{Koba2016} with its width representing the experimental error bars. Focussing at this figure we can discard the possibility of valence neutron being in the $1d_{3/2}$ or $1f_{7/2}$ orbital and hence, the possibility of $J^{\pi}=$ $3/2^+$ or $7/2^+$ for the ground state of $^{29}$Ne can be ruled out. Therefore, on the basis of single reaction observable i.e. the one-neutron removal cross section, we can restrict ourselves to $1/2^+$ and $3/2^-$ as the possible ground state $J^{\pi}$.


It is now interesting to estimate the range of $\beta_2$ that is consistent with the $s$- and $p$-wave ground state configuration of $^{29}$Ne at $S_n=0.96$\,MeV. This is illustrated in Fig.~\ref{1nx} for the $^{28}$Ne$(0^+) \otimes 2s_{1/2}\nu$ (dashed line) and  $^{28}$Ne$(0^+) \otimes 2p_{3/2}\nu$ (solid line). In case of $\ell = 0$, deformation has negligible effect on one-neutron removal cross section: this is also consistent with the observations for $^{31}$Ne and $^{37}$Mg which are in the same mass region \citep{Shubh2014, Shubh2015}. On the other hand, these calculations provide a limit for the possible quadrupole deformation value, i.e., $0.0$ $\leq \beta_2\leq 0.78$ for the $\ell = 1$ case. This corresponds to the $[330 1/2]$ state in terms of the Nilsson orbits under the $1p-2h [v(sd)^{-2} (fp)^{1}]$ scheme \cite{Koba2016, Hamamoto2004}. 

Of course, it is not possible to remove multiple discrepancies in the ground state of a nucleus with just one reaction observable as in the present case. This motivates us to calculate other exclusive observables like the parallel momentum distributions (PMD) and the relative energy spectra to put constraints on the spin-parity assignment.


\subsection{Parallel momentum distribution (PMD)}
We calculate the parallel momentum distribution of the core ($^{28}$Ne) in the elastic Coulomb breakup of $^{29}$Ne on $^{208}$Pb at $244$\,MeV/u beam energy for the possible ground state $J^{\pi}$ values of $^{29}$Ne. 
The values of full widths at half maxima (FWHM) in our calculations for $^{28}$Ne($0^+$)$ \otimes 2s_{1/2}\nu$ ($\ell = 0)$ and $^{28}$Ne  ($0^+$)$ \otimes 2p_{3/2}\nu$ ($\ell = 1$) configurations come out around $52$ and $82$\,MeV/c, respectively. For the other two configurations of $^{29}$Ne which have already been ruled out in Section~\ref{1n}, i.e., for $^{28}$Ne($0^+$)$ \otimes 1d_{3/2}\nu$ and $^{28}$Ne($0^+$)$ \otimes 1f_{7/2}\nu$, the FWHM is calculated as $155$\,MeV/c and $207$\,MeV/c, respectively. These values are comparable to those available in the literature. For example, the FWHM of the PMDs have also been calculated from the different $^{29}$Ne($J^\pi$) Shell-Model (SM) wave functions by the authors of Ref. \citep{Koba2016}. The widths of the distributions for removal of a neutron corresponding to $\ell=0, 1, 2$ and $3$ states were reported as $51.5$, $84.8$, $175.2$, and $281.7$\,MeV/c, respectively.

However, out of these values only the FWHM corresponding to $\ell= 1$ is close to the experimental value of $98(12)$\,MeV/c \citep{Koba2016}. This clearly favors the $J^{\pi} = 3/2^-$ for the ground state of $^{29}$Ne. Nevertheless, we also consider mixed configurations (involving excited states of the core) for $J^{\pi} = 3/2^-$ and $J^{\pi} = 1/2^+$ as was done in Ref. \citep{Koba2016}, adopting the same SM spectroscopic factors (SFs). In fact, in Ref. \citep{Koba2016}, it was found that the PMD data can be fitted only for $J^{\pi} = 3/2^-$, that too after considering the mixing of various configurations of the same parity. In our calculations as well, we get an FWHM of 93 MeV/c for $J^{\pi} = 3/2^-$ when we consider mixing of various states, while it is around 180 MeV/c for the $J^{\pi} = 1/2^+$. This, therefore, confirms the $J^{\pi} = 3/2^-$ assignment to the ground state of $^{29}$Ne and favors the findings of Ref. \citep{Koba2016}.

\begin{figure}[h!]
	\centering
	\resizebox{0.80\textwidth}{!}{\includegraphics[clip,trim=0cm 0cm 0cm 0cm]{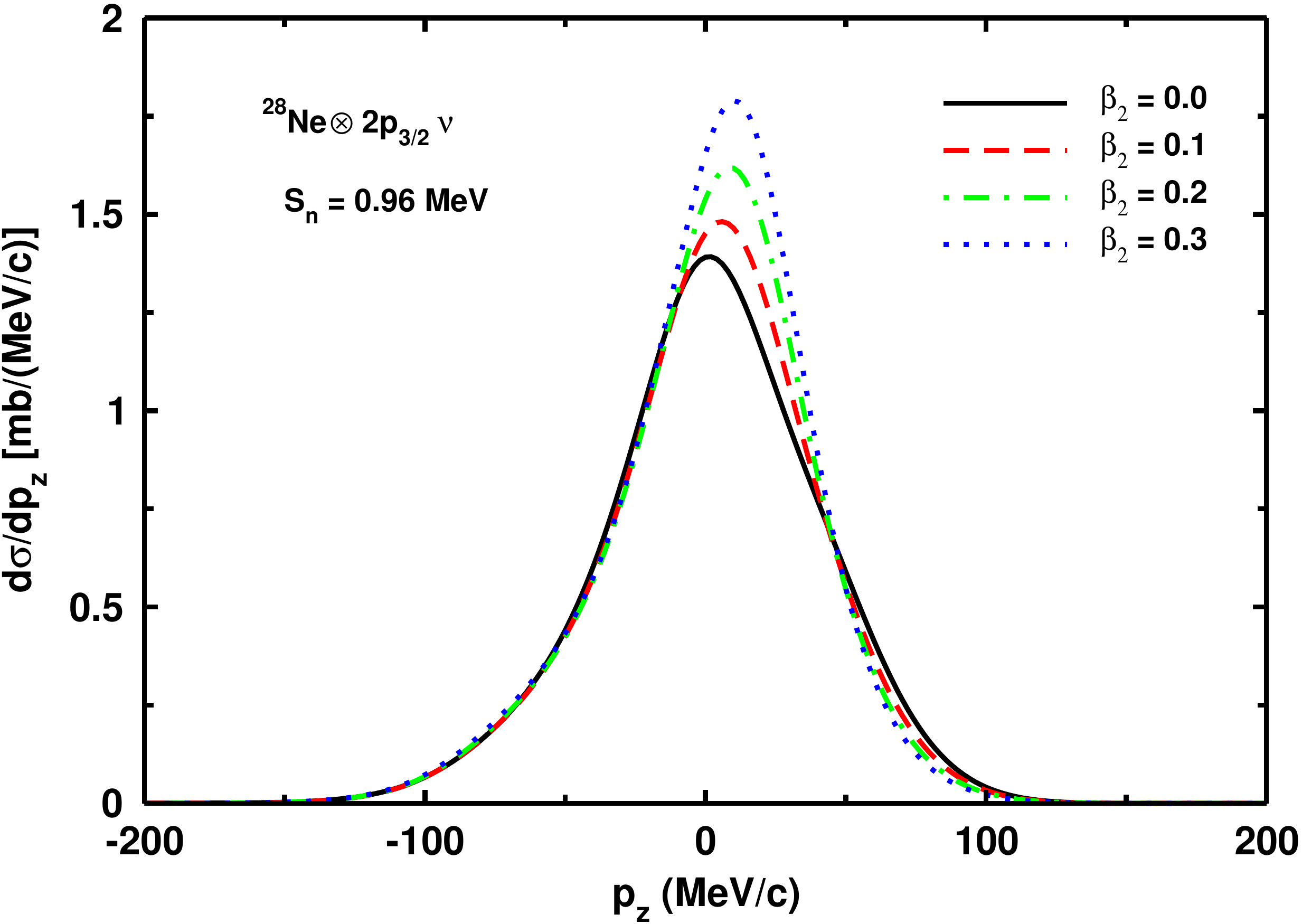}}
	\caption{The parallel momentum ($p_z$) distributions of $^{28}$Ne for the elastic Coulomb dissociation of $^{29}$Ne on $^{208}$Pb at $244$\,MeV/u in the rest frame of the projectile. The g.s. of $^{29}$Ne is accepted to be $^{28}$Ne$(0^+$)$ \otimes2p_{3/2}\nu$ with a neutron separation energy of $0.96$\,MeV. The solid, dashed, dash-dotted, and dotted lines correspond to $\beta_2=0.0$, $0.1$, $0.2$, and $0.3$, respectively. The distributions become narrower with increase in the quadrupole deformation. For details, see text.}
	\label{parl}
\end{figure}

Next we study the effect of deformation on the PMD and also try to further constrain the $\beta_2$ value for $^{29}$Ne in its ground state. As one can see the $^{28}$Ne($0^+$)$ \otimes 2p_{3/2}\nu$ component contributes around $\approx90\%$ to the total FWHM for the $J^{\pi} = 3/2^-$ ground state of $^{29}$Ne, therefore, for simplicity, we study the effect of deformation only considering this configuration of $^{29}$Ne. In Fig. \ref{parl}, we plot the PMD of the core in the Coulomb breakup $^{29}$Ne on Pb target at 244 MeV/u for four different values of $\beta_2$. One can see that the maximum effect of the deformation can be seen on the peak position where the peak height rises with the $\beta_2$. Subsequently, we study the effect of deformation on the FWHM values, which are given in Table \ref{fwhm} \footnote{The FWHM obtained with spectroscopic factor $0.7$ (as it can be seen from ref.\citep{Hamamoto2004}), for the $^{28}$Ne($0^+$)$ \otimes 2p_{3/2}\nu$ configuration, corresponding to deformations $\beta_2=0, 0.1$ and $0.2$ are $80, 74$ and $69$\,MeV/c, respectively. We observe that these values are not appreciably different with respect to those obtained with $C^2S=1$ given in Table \ref{fwhm}.}. It is clear from the table that with increase in the $\beta_2$, the FWHM decreases. Furthermore, going from $\beta_2$ = 0.0 to 0.2 the FWHM for the dominant $^{28}$Ne($0^+$)$ \otimes 2p_{3/2}\nu$ ground state component decreases by 13\%. Therefore, with the same proportion we can expect an FWHM of $\approx 81$ MeV/c at $\beta_2 = 0.2$, when considering all the SM suggested components of $J^{\pi} = 3/2^-$ (see Ref. \citep{Koba2016}). This value is already outside the limits of the experimental value of 98(12) MeV/c, and thus, prevents us from considering higher $\beta_2$ values. However, it must be mentioned that this is in slight contradiction to the Nilsson model predictions in Ref. \citep{Koba2016}, which suggest somewhat larger deformation values for the $^{29}$Ne ground state.


\begin{table}[htbp]
	\caption{FWHM of the parallel momentum distribution of $^{28}$Ne obtained in the elastic Coulomb breakup of $^{29}$Ne [$^{28}$Ne($0^+$)$ \otimes 2p_{3/2}\nu$] on a Pb target at $244$\,MeV/u beam energy. }
	\label{fwhm}
	\begin{center}
		\begin{tabular}{c c }
			\hline\hline
			$\beta_2$ &  FWHM (MeV/c)  \\
			\hline 
			\hline
			0.0 &  82 \\
			0.1 &  76 \\
			0.2 &  71 \\
			0.3 &  68 \\
			0.4 &  66 \\
			\hline\hline
		\end{tabular}
	\end{center}
\end{table}

Parallel momentum having a narrow distribution indicates a wider spatial distribution via Heisenberg's uncertainty principle. In fact, it is known that the FWHM of PMD of the charged core in the breakup of light halo nuclei like $^{11}$Be and $^{19}$C is around $44$\,MeV/c, whereas, for the corresponding case of stable nuclei, it is about $140$\,MeV/c \cite{Chatterjee2000}. Since $^{29}$Ne seems to most probably have a dominant $2p_{3/2}$ contribution to its ground state, an FWHM of $82$\,MeV/c manifests that under conditions of sphericity, it should be a moderately sized halo, with increase in deformation resulting in a larger spatial extension.

Although, with these two observables we are able to establish the ground state structure of $^{29}$Ne, but it is still important to discuss other reaction observables which are vital in manifesting the structure of loosely bound nuclei \cite{Chatterjee2018}. We now turn our attention to the relative energy spectra, which, apart from structural information, can also be used to obtain the binding energy \cite{Nagarajan2005,ber92}.

\subsection{Relative energy spectrum}

\begin{figure}[h]
	\resizebox{0.80\textwidth}{!}{\includegraphics[clip,trim=0cm 0cm 0cm 0cm]{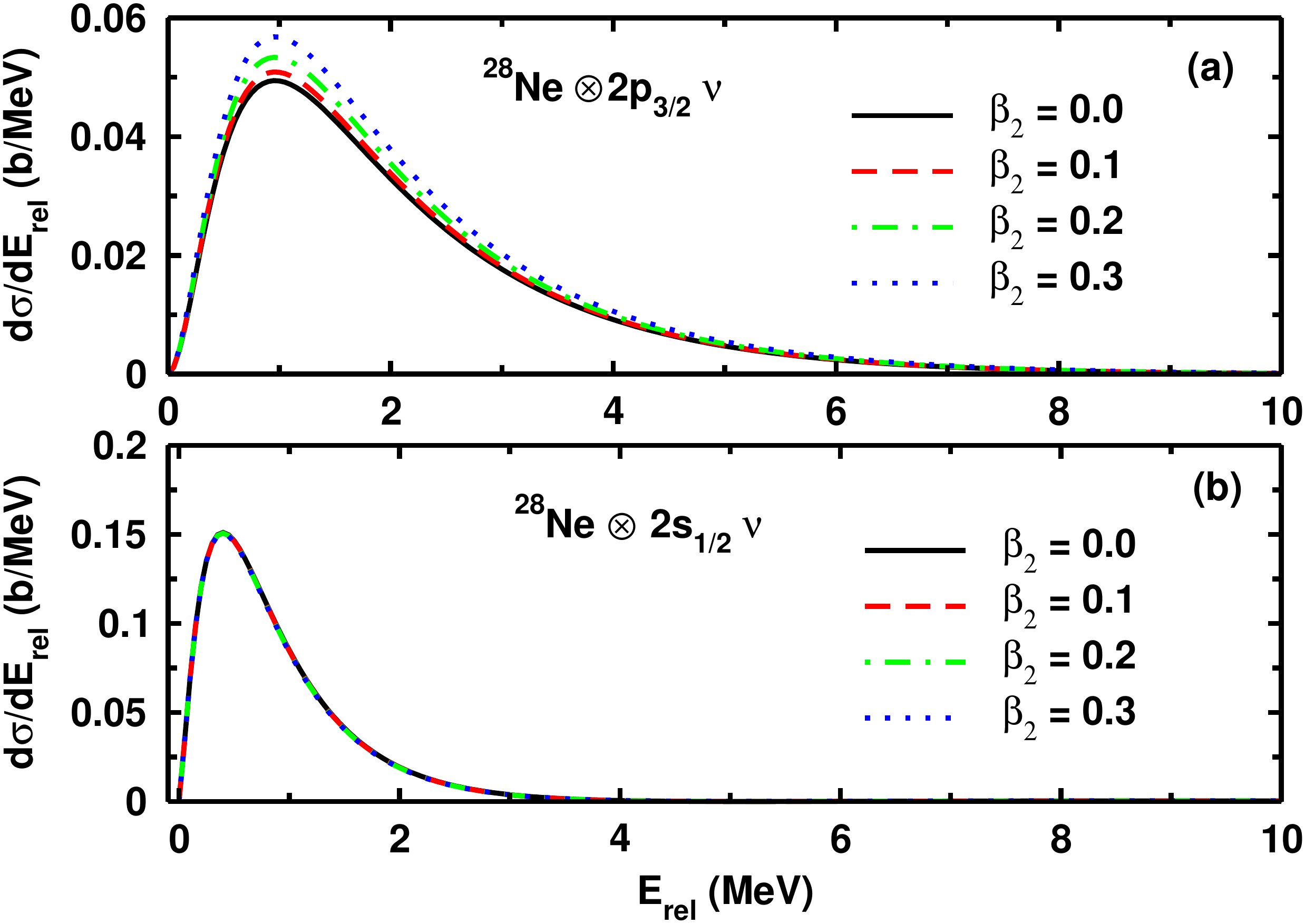}}
	\caption{The relative energy spectra of $^{29}$Ne breaking elastically on $^{208}$Pb at $244$\,MeV/u beam energy due to Coulomb dissociation. The deformation $\beta_2$ is varied from $0.0$ to $0.3$, for $S_n=0.96$\,MeV. (a) The g.s. of $^{29}$Ne is assumed to be formed by $2p_{3/2}$ contribution of the valence neutron. (b) The g.s. of $^{29}$Ne is assumed to be formed by $2s_{1/2}$ valence neutron contribution.}
	\label{relen}
\end{figure}

The relative energy spectra of $^{29}$Ne breaking up into $^{28}$Ne and a neutron on $^{208}$Pb at $244$\, MeV/u, is shown in Fig.~\ref{relen} (a), for $^{28}$Ne$(0^+) \otimes2p_{3/2}\nu$ configuration at $\beta_2=0.0$ (solid line), $0.1$ (dashed line), $0.2$ (dot-dashed line), and $0.3$ (dotted line),  respectively. The angular integration in the relative energy spectra has been carried out till the grazing angle of $1.25\degree$. It is clearly seen that the peak height is sensitive to the deformation, $\beta_2$. On the other hand, the relative energy spectra for $s$-state illustrated in Fig.~\ref{relen} (b), is not affected by any of the deformations. This is in line with the observations in Refs. \citep{Shubh2014, Gagan2016}, where the relative energy spectra for the breakup of $^{31}$Ne and $^{34}$Na was largely independent of the parameter $\beta_2$. However, the peak height, in $s$-wave configuration, is much larger than the one obtained for $^{28}$Ne$(0^+) \otimes2p_{3/2}\nu$ ground state configuration. Therefore, the experimental measurement of the relative energy spectra would be an independent check on the issue of the ground state of $^{29}$Ne being a $p$-wave or an $s$-wave.

\subsection{Neutron energy-angular distribution}

\begin{figure}[h]
	\centering
	\resizebox{0.80\textwidth}{!}{\includegraphics[clip,trim=0cm 0cm 0cm 0cm]{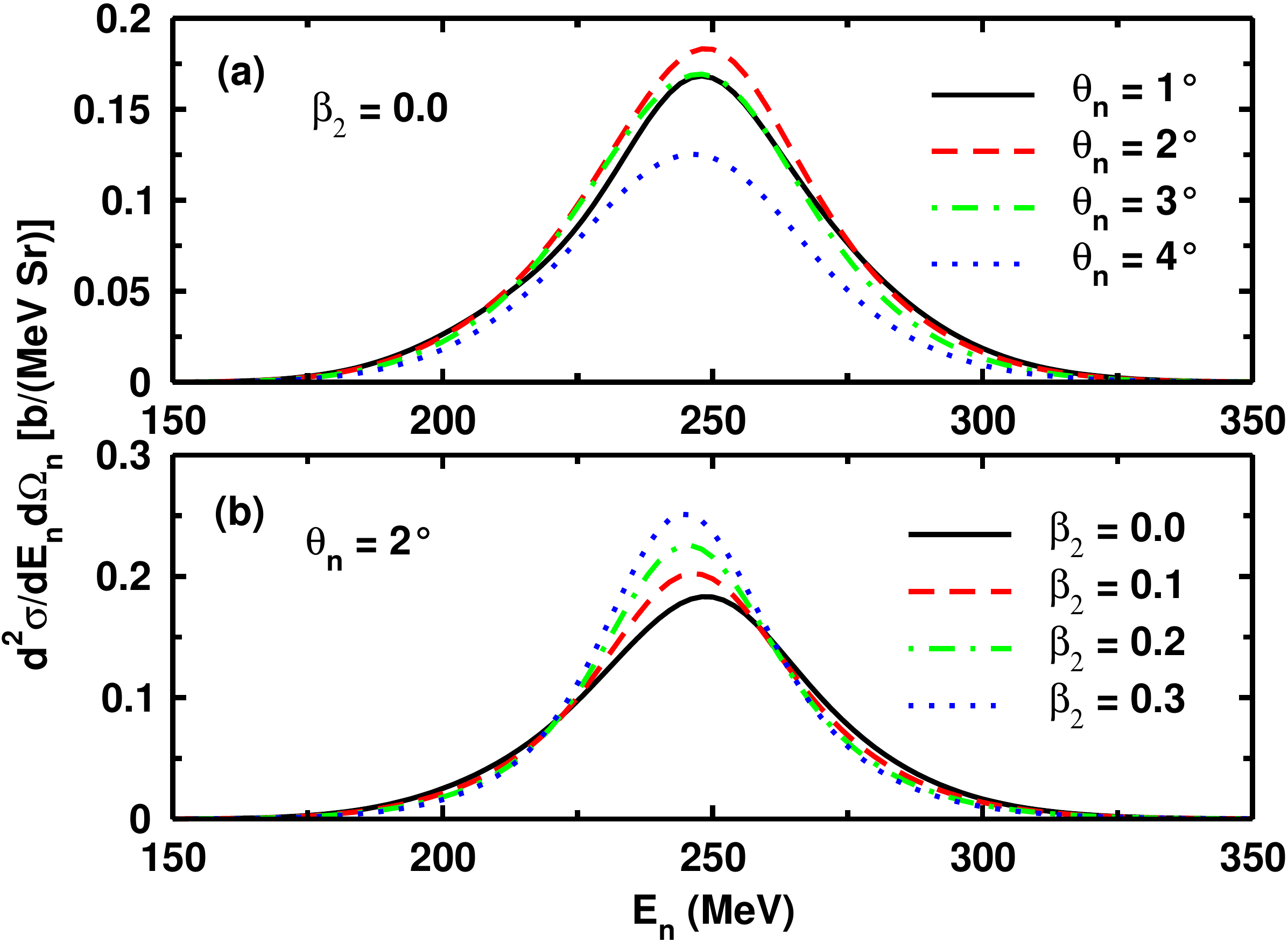}}
	\caption{(a) Neutron energy-angular distributions for the elastic Coulomb breakup of a spherical $^{29}$Ne on a $^{208}$Pb target at $244$\, MeV/u calculated for $S_n=0.96$\, MeV for projectile ground state $J^{\pi}=3/2^{-}$. Variation is shown with neutron angle, $\theta_n$. (b) The neutron energy-angular distributions for the same reaction with variation in the quadrupole deformation ($\beta_2$) parameter while the neutron angle was fixed at $\theta_n=2^{\circ}$.}
	\label{def} 
\end{figure}

In Fig.~\ref{def}(a), we show the neutron energy-angular distribution for the breakup of $^{29}$Ne on $^{208}$Pb target at ${244}$\, MeV/u for different neutron angles, $\theta_n=1^{\circ}$, $2^{\circ}$, $3^{\circ}$ and $4^{\circ}$. One can observe from Fig.~\ref{def}(a) that the peak position corresponding to different $\theta_n$'s is very close to the energy corresponding to the beam velocity, which indicates that there will be no post-acceleration in the final channel for the charged fragment $^{28}$Ne. This characteristic of the breakup of halo nuclei was previously seen in Ref. \citep{Shyam1992} and later also by the authors of Refs. \citep{Canto1993, Bertulani1995, Kido1996, Anne1994}. The absence of post-acceleration is due to the fact that the projectile breaks up at quite a large distance from the target \citep{Shyam1992}. 

It would now be interesting to study whether the peak position of the neutron energy-angular distribution would change with the deformation of the projectile. We address this question in Fig.~\ref{def}(b), where for a fixed neutron angle $\theta_{n}=2^{\circ}$, the neutron energy-angular distribution is illustrated for $\beta_2=0.0$ (solid line), $0.1$ (dashed line), $0.2$ (dot-dashed line) and $0.3$ (dotted line).

It is evident from this figure that there is an increase in cross section with the increase in $\beta_2$, similar to the case of PMD. The projectile deformation effects can be easily noticed near the peak positions of the distributions.  Similar to Fig.~\ref{def}(a), the peak positions in Fig.~\ref{def}(b) nearly coincide with the energy corresponding to beam velocity irrespective of the extent of the deformation, and this further conveys the absence of a post-acceleration of the charged fragment. Nevertheless, to rule out the possibility of post-acceleration effects it would be interesting to study the average momentum of the charged fragment in the breakup process.

\subsection{Average momentum of the charged fragment}
\begin{figure}[h!]
	\centering
	\resizebox{0.80\textwidth}{!}{\includegraphics[clip,trim=0cm 0cm 0cm 0cm]{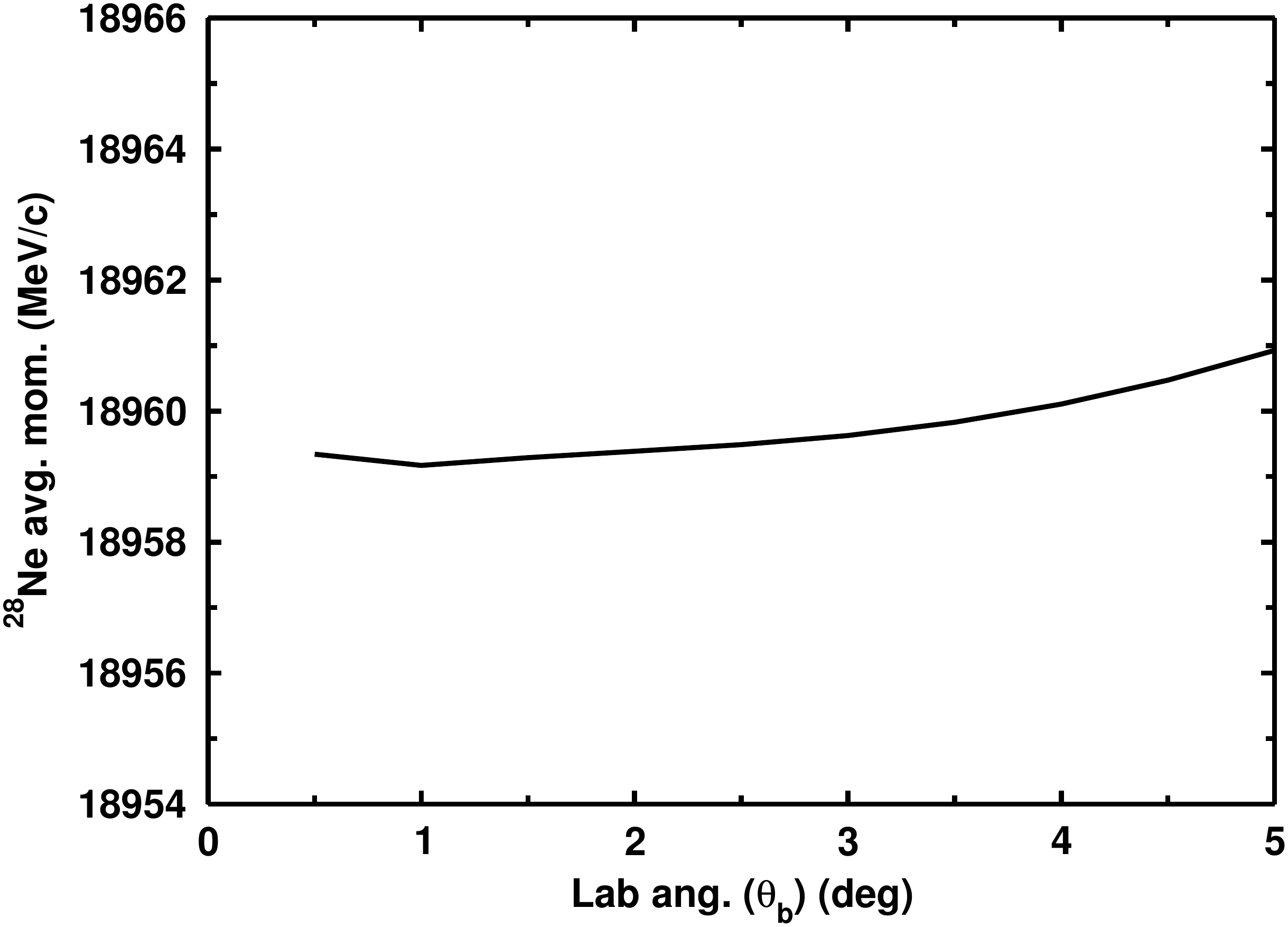}}
	\caption{Average momentum of the $^{28}$Ne fragment in the elastic Coulomb breakup of $^{29}$Ne on $^{208}$Pb at $244$\, MeV/u as a function of its scattering angle corresponding to the $^{28}$Ne($0^+) \otimes2p_{3/2}\nu$ ground state configuration.}
	\label{avg}
\end{figure}

It was predicted that post-acceleration effects would visible if the average momentum{\footnote{The average momentum for the charged core is defined as $\sum p_d \dfrac{d^2 \sigma}{dp_d d\Omega_d} \bigg /\sum \dfrac{d^2 \sigma}{dp_d d\Omega_d}$, with $p_d$ being the linear momentum of the core $d$.}} of the charged particle increases with the scattering angle \citep{Baur1992}, due to the fact that a larger scattering angle would entail a smaller impact parameter, which in turn would imply a larger Coulomb repulsion. However, no such increase is observed in the average momentum of the $^{28}$Ne fragment in the breakup of $^{29}$Ne on $^{208}$Pb target at $244$\, MeV/u, as shown in Fig.~\ref{avg}. This would give further credence to the fact that at the given beam energy, post-acceleration effects would not be present in the breakup of $^{29}$Ne.

\section{Conclusion}
\label{con}

We investigate the ground state structure of $^{29}$Ne breaking elastically on a $^{208}$Pb target at a beam energy of $244$\,MeV/u by analyzing several reaction observables. We employ the method of Coulomb dissociation within the post form finite-range distorted wave Born approximation theory. The upshot of FRDWBA is that the transition matrix can be separated into two parts: one containing the structure, which has the effects of possible projectile deformation, and the other comprising of the dynamics part of the reaction that can be solved analytically \citep{Nord1954}. FRDWBA is also advantageous over several first-order formalisms as it is an all-order theory encompassing the complete non-resonant continuum contribution from the resultant breakup fragments.

By comparing our calculations with the available data on the one-neutron removal cross section and FWHM in the parallel momentum distribution of the core in the breakup process, we find that the ground state of $^{29}$Ne is most likely consistent with a $J^{\pi}$ of $3/2^-$ with a dominant contribution from the $^{28}$Ne($0^+) \otimes2p_{3/2}\nu$ configuration. The FWHM in the PMD comes out around 82 MeV/c for this dominant configuration whereas this value increases to 93 MeV/c when we take into account the mixing of various SM suggested states which are adopted from Ref. \citep{Koba2016}. We are further able to set up an upper limit on $\beta_2$ which in our case is $\leq 0.2$.
An FWHM of $93$\,MeV/c comparable with the experimentally observed value of $98(12)$\,MeV/c,  directs us to the verdict that if $^{29}$Ne is a halo nucleus, it has to be moderate sized with a dominant $^{28}$Ne($0^+) \otimes2p_{3/2}\nu$ ground state configuration. It may be worth mentioning that the range of $\beta_2$ predicted in our calculation could have been slightly different, had we used the proper deformed wave function. However, this would not have modified the conclusion that $^{29}$Ne is a moderate halo since the variation of the FWHM with an with an adjusted spectroscopic factor would not be too discernible.

We also calculate other reaction observables like relative energy spectra and neutron energy angular distributions and study the effect of projectile deformation on them. 
The neutron energy-angular distribution and the average momentum calculations manifest that the breakup of $^{29}$Ne is free from post-acceleration effects, which is to be expected due to the low binding energy of the projectile and the high beam energy employed for the breakup reaction study. 
Nevertheless, more theoretical as well as experimental analyses are required to constrain the as yet uncertain structural parameters like the deformation, etc.

Roles of higher order processes, like in the case of electromagnetic responses, multistep breakup, nuclear and Coulomb-nuclear interference (CNI) components, etc., are also likely to manifest interesting aspects of this multifaceted subject \cite{SinghJ2020}. We intend to discuss these subjects in the near future and will be reported elsewhere.

\section{Acknowledgment}
This text presents results from research supported by the Scheme for Promotion of Academic and Research Collaboration (SPARC/2018-2019/P309/SL), Ministry of Education Govt. of India. Support from Ministry of Education grants, Govt. of India, to [M] and [MD] are gratefully acknowledged. [GS] acknowledges the Spanish Ministerio de Ciencia, Innovaci\'on y Universidades and FEDER funds under project FIS2017-88410-P and the European Union's Horizon 2020 research and innovation program under grant agreement No. 654002. [S] acknowledge the funding from the European Union’s Horizon 2020 research and innovation programme under the Marie Skłodowska-Curie grant agreement No. 801505.

\end{document}